\documentclass[reprint,prb]{revtex4-1}

\usepackage{amsmath}
\usepackage{graphicx}
\usepackage{mathrsfs}
\usepackage{bm}
\usepackage{verbatim}

\usepackage{hyperref}
\usepackage{color}
\definecolor{jacksonsPurple}{RGB}{45,48,145}
\hypersetup{colorlinks,
	linkcolor={jacksonsPurple},
	citecolor={jacksonsPurple},
	urlcolor={jacksonsPurple}
}

\newcommand{\UCLAPhysics}{Department of Physics and Astronomy, University of California at Los Angeles, \\ 405 Hilgard Avenue, Los Angeles, California 90095, USA}

\begin{document}
	
	\title{Efficient unitary method for simulation of driven quantum dot systems}
	
	\author{Spenser Talkington}
	\email{stalkington@ucla.edu}
	\affiliation{\UCLAPhysics}
	
	\author{HongWen Jiang}
	\affiliation{\UCLAPhysics}
	
	\date{\today}
	
	\begin{abstract}
		Density matrices evolved according the von Neumann equation are commonly used to simulate the dynamics of driven quantum systems.  However, computational methods using density matrices are often too slow to explore the large parameter spaces of solid state quantum systems.  Here we develop a unitary computation method to quickly perform simulations for closed quantum systems, where dissipation to the environment can be ignored.  We use three techniques to optimize simulations, apply them to six time-dependent pulses for a semiconductor quantum dot qubit system, and predict the dynamic evolutions. We compare computational times between our unitary method and the density matrix method for a variety of image sizes. As an example, we implement our unitary method for a realistic four-state system [Z. Shi, \textit{et al.}, \href{https://doi.org/10.1038/ncomms4020}{Nat. Commun. \textbf{5}, 3020 (2014)}], and find that it is over two orders of magnitude faster than the corresponding density matrix method implemented in the popular quantum simulation software QuTiP.
	\end{abstract}
	
	\maketitle
	
	\section{Introduction}
		The evolution of multi-state quantum systems with time-dependent driving is a fundamental problem in quantum information. While it is simple to specify an initial state, a Hamiltonian and its time dependence, determining the final state from these specifications is challenging. The standard method is to use density matrices \cite{korotkov1999continuous,stievater2001rabi,kiesslich2007noise,vznidarivc2011quantum,culcer2012valley,schoenfield2017coherent,shi2014fast}. While density matrix methods are robust and extensible, their numerical implementations are computationally expensive.
		
		Two research methodologies depend on the simulation of driven quantum systems. In the first, experiments are performed and their results are then backed by simulations. In the second, simulations are conducted and experiments seek to realize their behavior. Yet, both methodologies have challenges. In the first, the challenge is to determine the Hamiltonian that reproduces the experimental results. Finding this Hamiltonian requires a parameter space exploration. In the second, finding evolutions that warrant experimental study requires a parameter space exploration, and the experimental investigation may be challenging.
		
		The parameter spaces of solid state quantum systems are usually very large. For example, consider a charge based quantum dot qubit in silicon. An ideal qubit is a two-state system, yet in real solids such as silicon, realizations of qubits are more complex. The realization of a charge qubit involves the creation of two quantum dots, where the position encodes the quantum state \cite{lent1997device,loss1998quantum,hollenberg2004charge,petta2005coherent,ladd2010quantum,zwanenburg2013silicon}. Yet in real solids, there are additional pseudospin/valley states \cite{goswami2007controllable,culcer2012valley,hao2014electron}. In silicon, there are six valley states; considering the lowest two states turns the simulation of a qubit system into the simulation of a four-state system.
		
		Here we develop three techniques for efficient unitary evolution and present this method as a faster alternative to density matrix methods for the exploration of parameter spaces. Previous studies of unitary evolution and related methods have justified the decomposition of the unitary evolution operator into a product of exponentials of time-independent Hamiltonians \cite{suzuki1993improved,glaser1998unitary,white2004real,daley2004time,verstraete2004matrix,sakurai2014modern,blanes2017symplectic,bader2018exponential}. Meanwhile, J.C. Tremblay, \emph{et al}. preconditioned adaptive step sizes were introduced for Runge-Kutta solutions to the Schr\"odinger Equation in Ref.~\onlinecite{tremblay2004using}. In addition, we apply our method to the simulation of a realistic four-state system realized in a charge based quantum dot qubit system in a semiconductor by Z. Shi, \emph{et al}. in Ref.~\onlinecite{shi2014fast}.
		
		\subsection{Coherent Oscillations}
		Coherent oscillations in the probability of a state arise during time evolution as a result of quantum tunneling or spin precession. While coherent oscillations for simple systems such as the Larmor Precession of a spin-$1/2$ particle in a magnetic field are easy to visualize theoretically, the situation for multi-state systems is less clear.
		
		Consider a two-state system with eigenstates $|L\rangle$ and $|R\rangle$, corresponding to two quantum dots with potentials $V_L$ and $V_R$. Now consider a tunnel coupling of $\Delta$ between the two dots. The Hamiltonian for this system is then, with detuning $\epsilon$ defined as $\epsilon \equiv V_R - V_L$ \cite{petersson2010quantum,petersson2010charge,dovzhenko2011nonadiabatic}:
		\begin{align}
		H = \begin{pmatrix}
		\langle R|R\rangle & \langle R|L\rangle\\
		\langle L|R\rangle & \langle L|L\rangle\\
		\end{pmatrix}
		= \begin{pmatrix}
		\epsilon/2 & \Delta\\
		\Delta & -\epsilon/2\\
		\end{pmatrix}
		\end{align}
		
\begin{figure}
			\includegraphics[width=\columnwidth]{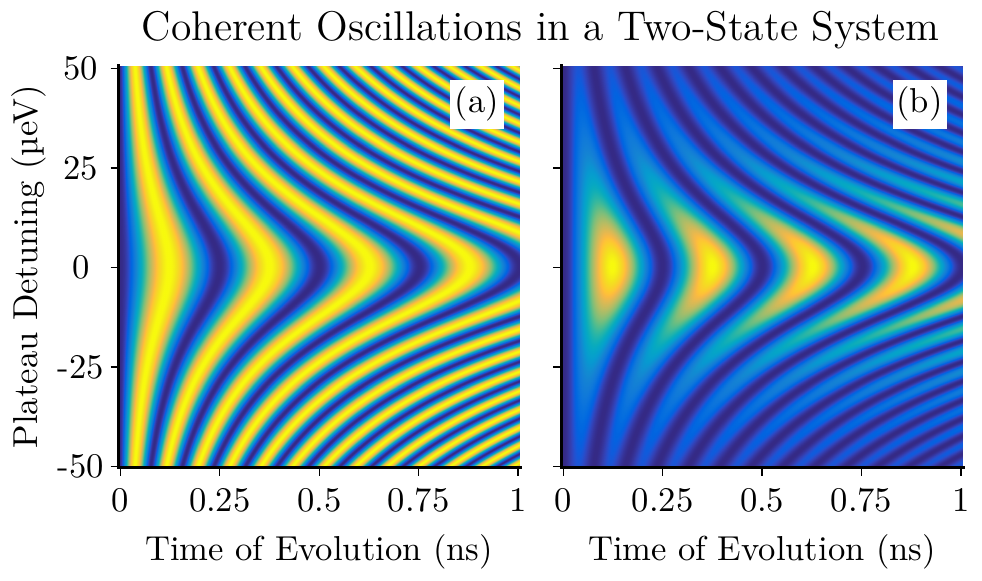}
			\caption{Coherent oscillations of $\langle R\rangle$ in the time evolution of a two-state system beginning with $|\psi(0)\rangle = |L\rangle$. Tunneling is set as $\Delta=4\pi\hbar$, where $\hbar = 0.658~\mathrm{\mu eV~ns}$. Dark is $\langle R\rangle=0$ and light is $\langle R\rangle=1$. (a) $\sin^2(\omega t)$ is an intuitive first approximation to the coherent oscillations. While an exact solution is possible here, its derivation is somewhat tedious and its form is illuminating. (b) Unitary evolution reveals the finer structure of the coherent oscillations.}
			\label{fig:chevron}
		\end{figure}

		With unitary evolution at constant $\Delta$ and $\epsilon$, this leads to coherent oscillations with a frequency of $\omega = E/\hbar =\sqrt{\Delta^2+\epsilon^2/4}/\hbar$. Such coherent oscillations are most clearly visualized on a image of probability amplitude with detuning plateau and time evolved. In Figure~\ref{fig:chevron}, oscillations occur at a frequency $\omega$, and peak amplitudes depend on the detuning.
		\footnote{This two-state constant Hamiltonian evolution is Larmor Precession. In terms of the Pauli spin matrices this Hamiltonian is $H = \Delta \sigma_x + (\epsilon/2) \sigma_z$, which for $\bm{B} = (-\Delta/\gamma,0,-\epsilon/2\gamma)^T$ is equivalent in form to the Hamiltonian for a spin-$1/2$ particle in a magnetic field: $H=-\gamma \bm{B}\cdot\bm{\sigma}$.}
	
		Images of coherent oscillations can be determined for real systems. Experimentally, this involves fabricating a system of quantum dots and electrodes. The electrodes are then connected to a pulse generator which sets electric potentials on the quantum dots \cite{hayashi2003coherent,petersson2010quantum,kim2014quantum}.
		
		In modeling these systems, the detuning is time dependent and experimentally relevant time dependencies may be considered---see Fig.~\ref{fig:pulses} for some examples. For example, a perfectly square pulse cannot be experimentally generated, so a trapezoid pulse approximates the finite rise and fall time characteristics of the pulse generator. Additionally, dots may have on-site splittings.
		
		These factors result in new physics  \cite{goswami2007controllable,culcer2012valley,hao2014electron,shi2014fast,schoenfield2017coherent}, and necessitate numerical solutions to determine accurately model the time evolution. Computationally, states may be evolved using density matrix methods, or, if there is no dissipation, unitary evolution methods. We do this by selecting an initial state, evolving through the pulse, and conducting post-pulse averaging for each pulse duration and detuning amplitude in some range. The final state is taken to be the average of the state at the minimum detuning over a readout time following the pulse.
		
		\subsection{Density Matrix Evolution}
		
		Currently, density matrix methods are the standard method used to model the time evolution of quantum information systems \cite{korotkov1999continuous,stievater2001rabi,kiesslich2007noise,vznidarivc2011quantum,culcer2012valley,shi2014fast, schoenfield2017coherent}. While these methods are accurate, and can create images from known parameters, they are too slow to explore the parameter space of real quantum dot systems.
		
		Density matrices, $\rho$, encode state information in a matrix of the form $\rho=\sum p_j |\psi_j\rangle\langle\psi_j|$, for positive real $p_j$ and $\mathrm{tr}(\rho)=1$. Expectation values of observables, $Q$, are computed as $\langle Q\rangle = \mathrm{tr}(Q\rho)$. Density matrices evolve according to a differential equation such as the von Neumann Equation \cite{chen2006quantum}:
		\begin{align}
		\dot{\rho} = -\frac{i}{\hbar} [H,\rho]
		\end{align}
		
		Density matrix methods then implement numeric solutions to the differential equation. Two such methods are the $4$th Order Runge-Kutta Method (RK4), and the Adams Linear Multistep Method (as is used in the QuTiP Library \cite{johansson2012qutip,johansson2013qutip}).
		
		\subsection{Unitary Evolution}
		As an alternative, we present unitary evolution. While unitary evolution cannot describe dissipation or coupling to the environment, the methods are, after optimization, fast and effective ways to approximate state evolution and to develop a theoretical understanding of real systems.
		
		Here we develop three techniques to optimize unitary evolution methods for determining images of coherent oscillations in the time evolution of quantum-dot systems driven by detuning pulses. When particle number $\langle \psi | \psi \rangle$ is conserved, quantum states evolve with a unitary time-evolution operator:
		\begin{align}
		\left|\psi(t)\right\rangle = \mathscr{U}(t,0) \left|\psi(0)\right\rangle
		\end{align}
		
		If we know the time-evolution unitary $\mathscr{U}(t,0)$ and the initial state $\left| \psi(0)\right\rangle$, then we are done. Yet it is often challenging to find $\mathscr{U}(t,0)$. In particular, if the state is a mixed state which is entangled with the environment or if the Hamiltonian has a probabilistic dependence, then it is hard to find the corresponding unitary, and it is simpler to use the density matrix formalism for time evolution. Otherwise, so long as Hamiltonians of different times commute the time-evolution unitary may be found using the formula \cite{sakurai2014modern}:
		\begin{align}
		\mathscr{U}(t,0) = \exp\left[-\frac{i}{\hbar} \int_0^t d\bar{t} \ H(\bar{t})\right]
		\end{align}
		
		Now, we may approximate the integral by expressing the unitary evolution as product of $N$ steps of constant $H$ spaced at $\tau_n \in \{\tau_0,\tau_1,\dots,\tau_N\} \subset (0,t)$, $\tau_0 = 0$, $\tau_N = t$:
		\begin{align}
		\mathscr{U}(t,0) &= \mathscr{U}\left(\tau_N,\tau_{N-1}\right) \dots \mathscr{U}\left(\tau_2,\tau_1\right) \mathscr{U}\left(\tau_1,\tau_0\right)\\
		&\approx \prod_{n=1}^N \exp\left[-\frac{i}{\hbar} (\tau_n-\tau_{n-1}) \, H\!\left(\tau_n\right)\right]
		\label{eq:decomposition}
		\end{align}
		
		Carefully selecting the time intervals $\tau_n$, and how the product is evaluated and stored leads to substantial improvements in computational time. Here $N$ is the total number of time steps in the simulation of a system, and we fix $\mathrm{num}$ as the image size $\mathrm{num}\times \mathrm{num}$.
		
		In particular, in the diagonal basis $\{|a_j^{(n)}\rangle\}$ for energies $\{E_j^{(n)}\}$ at each times $\tau_n$, the unitary operator is the time-ordered matrix product of the sum over the states:
		\begin{align}
		\mathscr{U}(t,0)
		&\approx \prod_{n=1}^N \sum_{j=1}^\mathrm{dim} \exp\left[-\frac{i}{\hbar} (\tau_n-\tau_{n-1}) E_j^{(n)}\right] |a_j^{(n)}\rangle\langle a_j^{(n)}|
		\label{eq:exact_diagonalization}
		\end{align}
		
		\subsection{Charge Based Semiconductor Quantum Dot Qubits}
		Multi-state quantum systems are commonly realized in semiconductors. Of particular interest are silicon qubits, where quantum dots encode state information. State information may be stored as spin, charge, or on-site valley splitting states. In this paper, we consider examples of charge qubits composed of quantum dots with valley splittings. The coherent oscillations of the electronic states in quantum dots are of key interest for the encoding and measurement of quantum information \cite{petta2004manipulation,oi2005robust,petersson2010charge,schoenfield2017coherent}.
		
		\subsection{This Paper}
		This paper aims to answer the question: ``how can we computationally explore the dynamics of driven quantum systems in the large parameter space of solid state quantum systems?" Our answer is: unitary evolution with optimizations for realistic driving pulse characteristics.
		
		In Section~\ref{sec:optimization}, we develop optimization techniques for unitary evolution. These optimization techniques depend on assuming the form of the time dependence, but these assumptions are usually experimentally applicable.
		
		In Section~\ref{sec:pulses}, we present six pulses: square, trapezoidal, ramp, sine, arc, and noise, discuss optimization techniques for unitary evolution methods with these detuning pulses.
		
		In Section~\ref{sec:comparison}, we compare the computational times and time-complexities of the optimized unitary methods with density matrix methods implemented in QuTiP for the three state system developed by Schoenfield, \textit{et al}. in Ref.~\onlinecite{schoenfield2017coherent}.
		
		In Section~\ref{sec:application}, we apply an optimized unitary method for a trapezoidal pulse to the realistic four-state system developed by Shi, \textit{et al}. in Ref.~\onlinecite{shi2014fast}.
		
	\section{Optimization Techniques}
	\label{sec:optimization}
		A simple implementation of unitary evolution for images of coherent oscillations in state's evolution is to use two nested \verb|for| loops: one over the detunings and one over the pulse times, with a fixed step size for unitary evolution. However, this is not particularly efficient. If we assume that the only time dependence of the Hamiltonian comes from the detuning, then we may optimize the evolution method. In particular, we develop three optimization techniques: evolution over constant intervals, extraction of repeated features, and extension of prior computations.
		
		\subsection{Constant time dependence}\label{sec:swctd}
			For unitary evolution, with the diagonal-time-dependence Hamiltonians that arise in quantum dot systems, no accuracy is gained by subdividing intervals that have a time-independent Hamiltonian. Therefore a unitary evolution method may be preconditioned to evaluate the evolution of such intervals in one step:
			\begin{align}
			\mathscr{U}_\mathrm{interval} = \exp\left[-\frac{i}{\hbar} t_\mathrm{interval} H_\mathrm{interval}\right]
			\end{align}
			
			This is useful for square pulses, nearly square pulses, and pulses with steps.
		
		\subsection{Repeated features}\label{sec:rf}
			Repeated features need only be calculated once. In particular, with the time-ordering operator $\mathcal{T}$:
			\begin{align}
			\mathscr{U}_\mathrm{total} = \mathcal{T} \left( \mathscr{U}_\mathrm{features}\mathscr{U}_\mathrm{other}\right)
			\end{align}
			
			This is useful for experimentally generated pulses which have fixed rise and fall times that are independent of pulse times, and other pulses with features that are repeated.
			
		\subsection{Extension of previous computations}\label{sec:eopc}
			If a new pulse is an extension of an old pulse, then by the separability of Eq.~\ref{eq:decomposition}:
			\begin{align}
			\mathscr{U}_\mathrm{total} = \mathscr{U}_\mathrm{new} \mathscr{U}_\mathrm{old}
			\end{align}
			
			This is useful for all pulses that progress uniformly, or have sections in between repeated features that progress uniformly.
		
		\subsection{Combining techniques}
			Which techniques are appropriate for a given time dependence must be considered carefully, and it should be noted that multiple techniques may be used for a single detuning pulse.
	
	\section{Pulses}\label{sec:pulses}
\begin{figure}
			\includegraphics[width=\columnwidth]{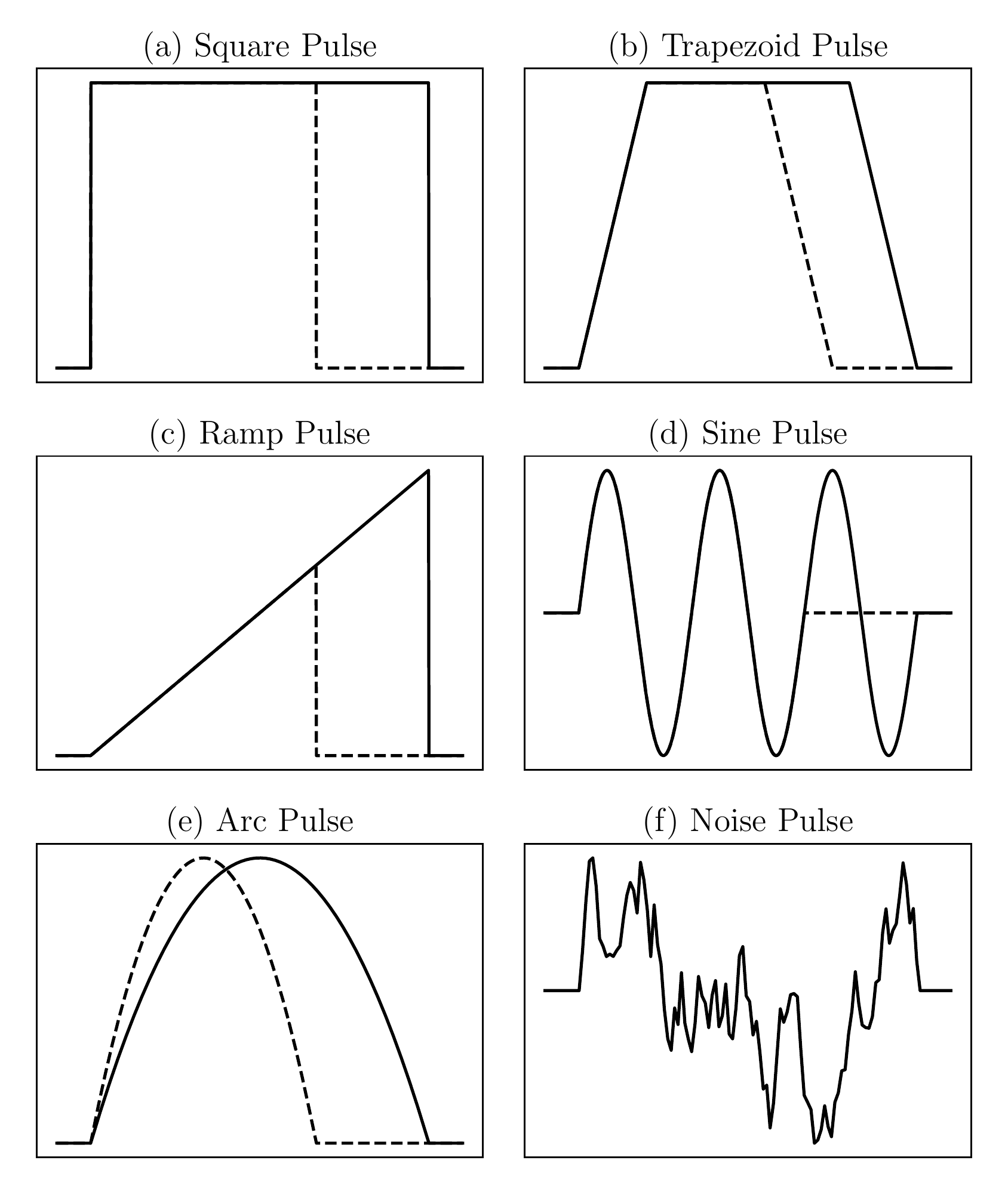}
			\caption{Plot of detuning versus time for six detuning pulses. In (a-e) a shorter pulse time is plotted with a dashed line and a longer pulse time is plotted with a solid line..
				(a) In pulses with intervals that are time independent, the unitary for that interval may be computed in one step.
				(b) In pulses with constant features like rises/falls (separated by a variable plateau), the features only need to be computed once.
				(c-d) In pulses where extending pulse time appends new detunings, it is only necessary to compute the unitary for the new terms and multiply.
				(e) In pulses where extending time is a dilation, the unitary method cannot be optimized by the techniques used in (a-d).
				(f) In pulses where the detuning at a given time is not known a priori, the unitary method cannot be optimized.}
			\label{fig:pulses}
		\end{figure}

		Here, we present six detuning pulse patterns, as seen in Fig.~\ref{fig:pulses}, and describe methods for optimizing unitary evolution with such time dependencies, and discuss computational times. See Appendix~\ref{sec:appendix_pseudocode} for pseudocode implementations of these methods.
		
		\subsection{Square}
			Consider a pulse that is a constant detuning, as in Fig.~\ref{fig:pulses}(a). An expression for a pulse from time $t=0$ to $t=t_p$ with detunings $\epsilon_\mathrm{min}$ and $\epsilon_\mathrm{max}$ is:
			\begin{align}
			\epsilon(t) =
			\begin{cases}
			\epsilon_\mathrm{min} & t\leq 0\\
			\epsilon_\mathrm{max} & 0<t\leq t_p\\
			\epsilon_\mathrm{min} & t_p<t\\
			\end{cases}
			\end{align}
			
			For this pulse, the unitary for evolution may be calculated in one step as in Sec.~\ref{sec:swctd}.
			
		\subsection{Trapezoid}
			A more experimentally realistic pulse is a trapezoid pulse with fixed rise and fall time as in Fig.~\ref{fig:pulses}(b). An expression for a pulse from $t=0$ to $t=t_p$ with detunings from $\epsilon_\mathrm{min}$ and $\epsilon_\mathrm{max}$ is:
			{\small
			\begin{align}
			\epsilon(t) = 
			\begin{cases}
			\epsilon_\mathrm{min}& t<0\\
			\epsilon_\mathrm{min}+(\epsilon_\mathrm{max}-\epsilon_\mathrm{min})t/t_r& 0<t<t_r\\
			\epsilon_\mathrm{max}& t_r<t<t_p-t_f\\
			\epsilon_\mathrm{max}-(\epsilon_\mathrm{max}-\epsilon_\mathrm{min})[t-(t_p-t_f)]/t_f& t_p-t_f<t<t_p\\
			\epsilon_\mathrm{min}& t_p<t
			\end{cases}
			\end{align}}
			
			For this pulse, the unitary for evolution may be broken into three sections:
			\begin{align}
			\mathscr{U}_\mathrm{total} = \mathscr{U}_\mathrm{fall} \mathscr{U}_\mathrm{plateau} \mathscr{U}_\mathrm{rise}
			\end{align}
			If the rise and fall unitaries are calculated once per pulse height as in Sec.~\ref{sec:rf}, and the plateau is calculated in one step as in Sec.~\ref{sec:swctd}.
		
		\subsection{Ramp}
			A pulse which progresses uniformly from $t=0$ to $t=t_{p}$ with pulse heights from $\epsilon_\mathrm{min}$ and $\epsilon_\mathrm{max}$, with a discontinuity at $t=t_p$, as in Fig.~\ref{fig:pulses}(c), is:
			\begin{align}
			\epsilon(t) =
			\begin{cases}
			\epsilon_\mathrm{min}&t\leq0\\
			\epsilon_\mathrm{min} + (\epsilon_\mathrm{max}-\epsilon_\mathrm{min})t/t_{p}&0\leq t \leq t_p\\
			\epsilon_\mathrm{min}&t_p < t
			\end{cases}
			\end{align}
			
			For this pulse, the unitary for evolution may be computed iteratively as in Sec.~\ref{sec:eopc}.
		
		\subsection{Sine}
			Another experimentally realistic pulse is a sine pulse. To ensure that the pulse is continuous, select angular frequency $\omega$ such that $\omega=j \pi/t_p$ for an integer $j$. However, we consider a sine wave with a discontinuity at $t=t_p$, as in Fig.~\ref{fig:pulses}(d). An expression for a pulse of frequency $\omega$ from $t=0$ to $t=t_p$ with detunings centered on $\epsilon_\mathrm{center}$ with amplitude $\epsilon_\mathrm{amp}$ is:
			\begin{align}
			\epsilon(t) =
			\begin{cases}
			\epsilon_\mathrm{center}&t<0\\
			\epsilon_\mathrm{center} + \epsilon_\mathrm{amp}\sin(\omega t)&0<t\leq t_p\\
			\epsilon_\mathrm{center}&t_p<t
			\end{cases}
			\end{align}
			
			For this pulse, the unitary for evolution may be computed iteratively as in Sec.~\ref{sec:eopc}. While in principle, it would be possible to store $\mathscr{U}_\mathrm{period}$ and decompose:
			\begin{align}
			\mathscr{U}_\mathrm{total} = \mathscr{U}_\mathrm{remainder}(\mathscr{U}_\mathrm{period})^\mathrm{periods}
			\end{align}
			
			This method is discouraged because it yields only modest improvements over iterative time dependence, demands large memory usage, and is relatively difficult to implement and modify.
			
			Note that QuTiP can implement unitary evolution for periodic driving through its Floquet methods. We note that for a $200\times 200$ image with no averaging during readout that QuTiP's mesolve method took $6420$ sec, QuTiP's Floquet methods took $383$ sec, and the diagonal-basis unitary method took $3.6$ sec.
		
		\subsection{Arc}
			An experimentally unrealistic pulse for which the detunings have neither constant time dependence, repeated features, nor extension of previous computations is a parabolic arc from $t=0$ to $t=t_p$, with detunings from $\epsilon_\mathrm{min}$ to $\epsilon_\mathrm{max}$, is as in Fig.~\ref{fig:pulses}(e), and:
			\begin{align}
			\epsilon(t) =
			\begin{cases}
			\epsilon_\mathrm{min} & t\leq 0\\
			\epsilon_\mathrm{min} + (\epsilon_\mathrm{max}-\epsilon_\mathrm{min}) \left(\frac{2}{t_p}\right)^2 t(t_p-t) & 0\leq t \leq t_p\\
			\epsilon_\mathrm{min} & t_p<t
			\end{cases}
			\end{align}
			
			One might imagine that the computation could be optimized by $\mathscr{U}_\mathrm{new} = (\mathscr{U}_\mathrm{old})^{t_\mathrm{new}/t_\mathrm{old}}$ as for scalars, yet we have non-commuting matrices: $[H_\mathrm{new},H_\mathrm{old}]\neq 0$, so this method does not work. This type of pulse is therefore an example of the worst-case scenario for the unitary evolution methods.
			
		\subsection{Noise}
			Consider a pulse composed entirely of noise. One such pulse is shown in Fig.~\ref{fig:pulses}(f). Specifying $\epsilon(0)$, the range of permissible noise $\epsilon \in (-a,a)$, and a roughness factor $r$, the pulse is for $x = \mathrm{random}(-a,a)$:
			\begin{align}
			\epsilon(t) = \begin{cases}
			\epsilon(t-\tau) + rx & \text{default}  \\
			\epsilon(t-\tau) - rx & \text{if default gives } \epsilon(t) \notin (-a,a)
			\end{cases}
			\end{align}
			
			This evolution has no constant time dependence, and no repeated features, that we can know a priori. Such a pulse could be defined to have extension of previous computations as in Sec.~\ref{sec:eopc}.
		
	\section{Comparison of Methods}
	\label{sec:comparison}
		Here we compare the speed of density matrix and optimized unitary methods for determining image arrays of the time evolution. In particular, we consider the evolution of a two quantum dot system with valley splitting on the left dot: $|L\rangle\to|L_1\rangle,|L_2\rangle$ (a three-state system) subject to a trapezoidal detuning pulse as developed by Schoenfield \textit{et al}. in Ref.~\onlinecite{schoenfield2017coherent}. The Hamiltonian is:
		\begin{align}
		H = \begin{pmatrix}
		\epsilon/2&\Delta_1&\Delta_2\\
		\Delta_1&-\epsilon/2&0\\
		\Delta_2&0&-\epsilon/2+\delta
		\end{pmatrix}
		\label{eq:threeStateHamiltonian}
		\end{align}
		
		Parameters used for comparison are: fixed rise and fall times of $100~\mathrm{ps}$, time at detuning plateau of $0$ to $3~\mathrm{ns}$, detunings of $-200$ to $1200~\mathrm{\mu eV}$, and $\Delta_1=26.5~\mathrm{\mu eV}$, $\Delta_2=56.2~\mathrm{\mu eV}$, $\delta=23.0~\mathrm{\mu eV}$. Following pulses, all systems were evolved at readout for $1~\mathrm{ns}$, and then the state measurement was averaged over that $1~\mathrm{ns}$.
		
\begin{figure}
			\includegraphics[width=\columnwidth]{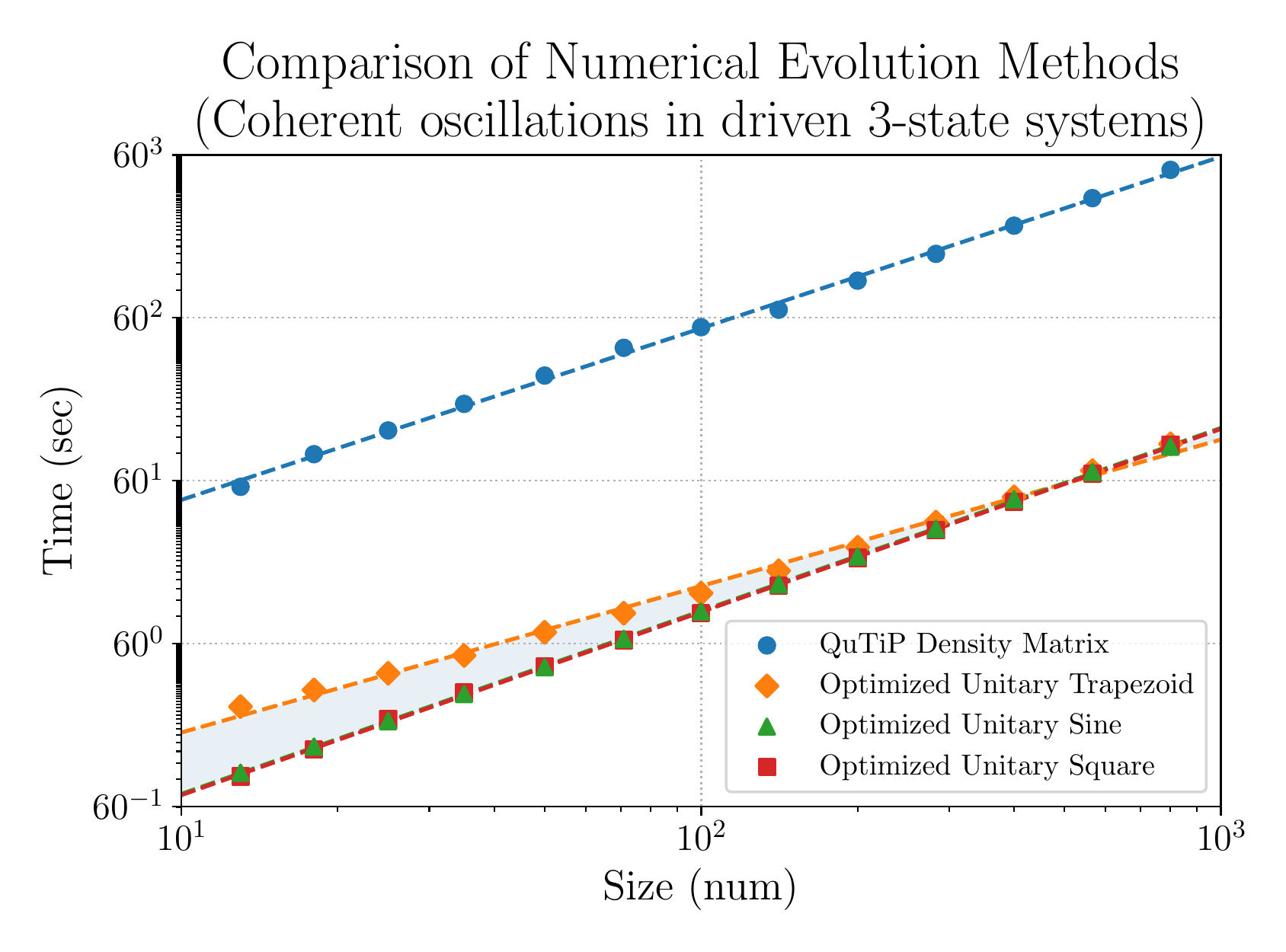}
			\caption{Computational times for determining images of coherent oscillations.
			At $\mathrm{num}=100$, the optimized unitary methods for sinusoidal, trapezoidal, and square pulses are between $800$ and $1300$ times faster than the QuTiP method. Note: $y$-axis scale is base-$60$.}
			\label{fig:compare}
		\end{figure}

		For a speed test, we selected values of the image size $\mathrm{num}$. For each size, we used (1) the QuTiP method to generate a $\mathrm{num}\times \mathrm{num}$ image of coherent oscillations with a trapezoidal detuning pulse, followed by evolution and pulse averaging at $\epsilon_\mathrm{min}$ (2) optimized unitary methods to generate $\mathrm{num}\times \mathrm{num}$ images of coherent oscillations with a trapezoidal, sinusoidal, and square detuning pulse, each followed by evolution and pulse averaging at $\epsilon_\mathrm{min}$. We timed each of the computations and plot the results in Fig.~\ref{fig:compare}. For comparison, computations were completed using Python on a single core of an \verb|Intel i7-7700HQ| processor. For QuTiP, the reporting step size was $10^{-10}~\mathrm{sec}$, and for the optimized unitary methods, the step size was $10^{-12}~\mathrm{sec}$ (which introduced errors of less than 1.5\%). See Appendix~\ref{sec:appendix_accuracy} and Appendix~\ref{sec:appendix_further} for details on these number and the accuracy and error bounds with step size.
		
		In Fig.~\ref{fig:compare}, we observe that while all solutions are computationally intensive for large systems, the optimized unitary method is much faster than the density matrix method at the same size. This is because step sizes and distribution of computations are more flexible for unitary methods than for solving solving differential equations.
		
		The best-fit lines to the data in Fig.~\ref{fig:compare} have different slopes. The slopes are the time complexity of the computational methods. The time complexities for the QuTiP method, and optimized unitary trapezoid, sine, and square methods are $O(\mathrm{num}^{1.87})$, $O(\mathrm{num}^{1.60})$, $O(\mathrm{num}^{2.00})$, and $O(\mathrm{num}^{2.00})$ respectively.
		
	\section{Application to a four-state system}
	\label{sec:application}
		While exploring the parameter space using density matrices is infeasible (a $4\times4$ hermitian matrix has $10$ free parameters, and determining one clear image ($\mathrm{num}\geq 100$) takes roughly one hour in QuTiP), such an investigation is possible using optimized unitary methods.
		Note that using optimized unitary methods, three-state systems and four-state systems compute in the same time
		\footnote{We observe that using optimized unitary methods, $3\times3$ and $4\times4$ systems run in roughly the same time. In particular for a $100\times100$ image, the three-state Hamiltonian runs in $22.4~\mathrm{sec}$ and the four-state Hamiltonian runs in $23.8~\mathrm{sec}$. Meanwhile in QuTiP for a $20\times20$ image, the three-state Hamiltonian runs in $36.7~\mathrm{sec}$ and the four-state Hamiltonian runs in $63~\mathrm{sec}$. We note that QuTiP time seems to scale linearly with the matrix size, i.e. $(16/9)36.7=65$.}.
		For examples of four-state parameter space evolution, see the supplementary material.
		
		\subsection*{Realistic four-state Hamiltonian}
		
			A more realistic Hamiltonian, with six time-independent parameters is developed by Shi \textit{et al.} in Ref.~\onlinecite{shi2014fast}. The Hamiltonian is:
			\begin{align}
			H&=\begin{pmatrix}
			\epsilon/2&0&\Delta_1&-\Delta_2\\
			0&\epsilon/2+\delta_L&-\Delta_3&\Delta_4\\
			\Delta_1&-\Delta_3&-\epsilon/2&0\\
			-\Delta_2&\Delta_4&0&-\epsilon/2+\delta_R\\
			\end{pmatrix}
			\label{eq:fourStateHamiltonian}
			\end{align}
			
\begin{figure}
				\includegraphics[width=\columnwidth]{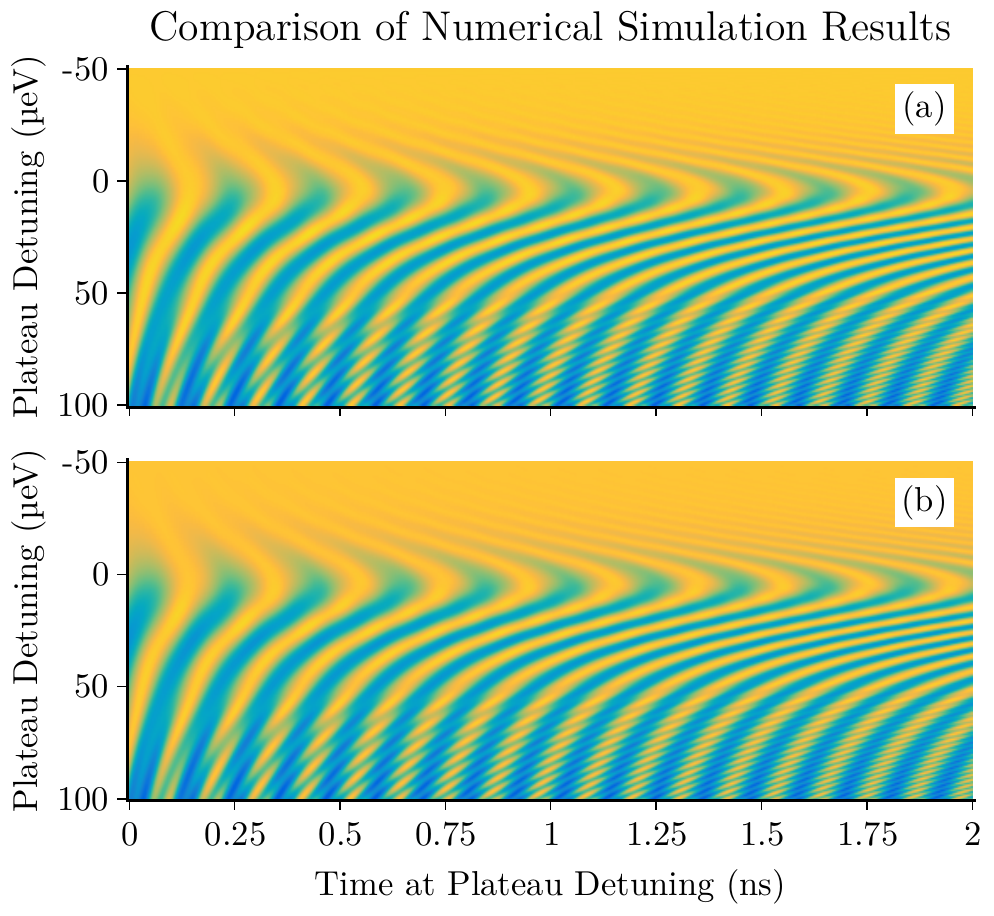}
				\caption{$500\times200$ images of coherent oscillations of $\langle L_1\rangle$ for a four-state system beginning in $|\psi(0)\rangle=|L_1\rangle$, driven by a trapezoidal pulse. The optimized unitary method for trapezoidal pulses is $546$ times faster than the QuTiP method. Tunnelings are as in the text below. (a) Density matrix methods using QuTiP took $5.92~\mathrm{hr}$ to determine this image. (b) The optimized unitary method took $39~\mathrm{sec}$ to determine this image.}
				\label{fig:fourState}
			\end{figure}

			Using the parameters in Ref.~\onlinecite{shi2014fast}, we demonstrate that the optimized unitary method can produce the same results as density matrix methods for a four-state Hamiltonian, assuming no decoherence or noise. The results are displayed in Fig.~\ref{fig:fourState}. We find that the optimized unitary method is two orders of magnitude faster than the density matrix method.
			
			The parameters used for four state evolution are the Hamiltonian in Eq.~\ref{eq:fourStateHamiltonian}, with $\Delta_1=10.83~\mathrm{\mu eV}$, $\Delta_2=14.47~\mathrm{\mu eV}$, $\Delta_3=19.02~\mathrm{\mu eV}$, $\Delta_4=6.82~\mathrm{\mu eV}$, $\delta_L=217.9~\mathrm{\mu eV}$, and $\delta_R=38.0~\mathrm{\mu eV}$, and a trapezoidal pulse with $t_r=0.118~\mathrm{ns}$ and $t_f=0.118~\mathrm{ns}$. The results for other detuning pulses may be found in the supplementary material.
			
	\section{Conclusions}
		We developed three optimization techniques for unitary evolution methods for determining images of coherent oscillations in driven quantum systems. These methods make assumptions on the form of the time dependence, here assumed to be a detuning pulse, in order to optimize computations. The methods act by taking advantage of experimentally relevant forms of detuning pulses: time-independent intervals of detuning, repeated detuning features, or subsequent detuning pulses that are extensions of previous detuning pulses. These assumptions allow the optimization of unitary evolution methods for pulses that are commonly realized experimentally including trapezoidal and sinusoidal detuning pulses. However, some pulses cannot be optimized using these techniques; for example, pulses which are dilations, and pulses which are not known a priori.
		
		We found that optimized unitary methods are much faster than the corresponding density matrix methods for all image sizes. In addition, we found that for the unitary methods, three-state systems and four-state systems evaluate in roughly the same time. We then applied an optimized unitary method and a QuTiP density matrix method to a realistic four-state system, and found that the optimized unitary method is over two orders of magnitude faster than the corresponding density matrix method.
		
		We believe that these techniques of unitary optimization can enable the computational exploration of the parameter spaces of coherent oscillations in quantum dot and multi-state systems with time dependencies such as the detuning pulses presented here. We expect such explorations to lead to a better theoretical understanding of driven multi-state systems, and they may lead to the discovery of new physical behaviors.
		
	\begin{acknowledgments}
		We gratefully acknowledge helpful discussions with N. E. Penthorn, J. D. Rooney, and T. J. Wilson surrounding this work.
		This work was supported by the UCLA Department of Physics and Astronomy, Summer REU Program and by U.S. ARO through Grant. No. W911NF1410346.
	\end{acknowledgments}
	
	\appendix
	\section{Accuracy of Methods}
	\label{sec:appendix_accuracy}
		For the comparison of computational methods, it is necessary to ensure that each method attains the same accuracy. Thus, we are interested in establishing a bound on the error of the optimized unitary methods, and finding computational parameters such that this error is less than a specified tolerance.
		
		To establish such a bound, we use the three-state Hamiltonian in Eq.~\ref{eq:threeStateHamiltonian} with the parameters in the main text, and consider a trapezoid pulse with a rise time of $0.1~\mathrm{ns}$, a fall time of $0.1~\mathrm{ns}$, and a plateau time of $0.8~\mathrm{ns}$, with the detunings $\epsilon_\mathrm{min}=-200$ and $\epsilon_\mathrm{plateau}=1200~\mathrm{\mu eV}$. The rise and fall in this pulse have the largest time dependence (diabaticity) of all pulses, and they therefore set a bound on the error.
		
\begin{figure}
			\includegraphics[width=\columnwidth]{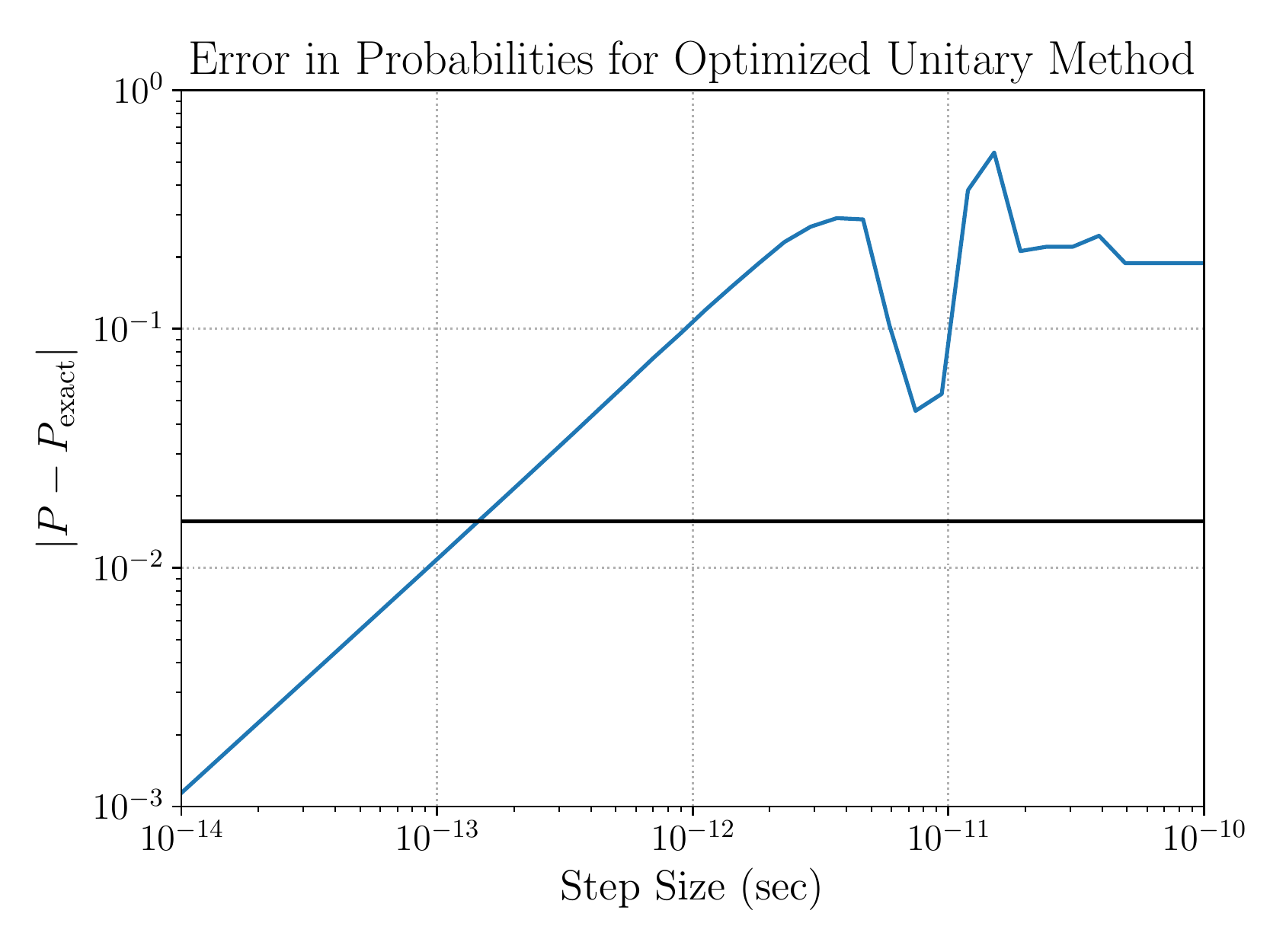}
			\caption{At a step size of $10^{-13}~\mathrm{sec}$, the optimized unitary method is accurate to within a tolerance of $1/64$. This plot is for a state evolving with a three-state Hamiltonian, and a pulse of the maximum diabaticity considered in this paper: from $-200~\mathrm{\mu eV}$ to $1200~\mathrm{\mu eV}$ in $0.1~\mathrm{ns}$.}
			\label{fig:accuracy}
		\end{figure}

		To set an ``exact" probability, we used a sixth-order Runge-Kutta density matrix method with a step size of $\tau_n = 10^{-15}~\mathrm{sec}$  on the rise/fall, and $\rho(0) = |R\rangle\langle R|$, to calculate the probability $P_\mathrm{exact}\equiv\langle R|\rho(1~\mathrm{ns})|R\rangle$ \cite{luther1968explicit}. From comparison with a fourth-order Runge-Kutta density matrix method, we believe $P_\mathrm{exact}$ to be accurate to within $10^{-10}$. Then, with the optimized unitary method, for a range of step sizes, we computed $P=\langle \psi(1~\mathrm{ns})|R\rangle\langle R|\psi(1~\mathrm{ns})\rangle$. We then took the magnitude of the difference. The results are plotted in Fig.~\ref{fig:accuracy}.
		
		Then we selected a tolerance on the accuracy. We used the Parula colormap to plot images of coherent oscillations, and this colormap has $64$ colors, so we specify the tolerance as $1/64$. In Fig.~\ref{fig:accuracy}, the optimized unitary method is accurate to within $1/64$ for a step size of $10^{-13}~\mathrm{sec}$.
		
		However, the maximum error is not the only reasonable measure of accuracy. We may instead consider the average error. For an image array $A$, the average error is $\mathrm{mean}|A-A_\mathrm{exact}|$. By computing $A_\mathrm{exact}$ with as above, and $A$ with a step size of $\tau_n = 10^{-12}~\mathrm{sec}$, we find that $\mathrm{mean}|A-A_\mathrm{exact}|<1/64$, so we use this step size.
		
		Note that in QuTiP, there is no direct control over step size. Instead, QuTiP uses adaptive step sizes to ensure an accuracy of $10^{-3}$ to $10^{-4}$, and reports expectation values at a series of time steps. Therefore we select the largest reporting step size that converges: $10^{-10}~\mathrm{sec}$.
	
	\section{Further Optimizations}
	\label{sec:appendix_further}
		The optimized unitary methods presented in this paper used fixed step sizes, $\tau_n$, for the time-dependent intervals of the Hamiltonian. However, it is possible to adapt the step size, either for each row of detuning, or for individual evolutions.
		
\begin{figure}
			\includegraphics[width=\columnwidth]{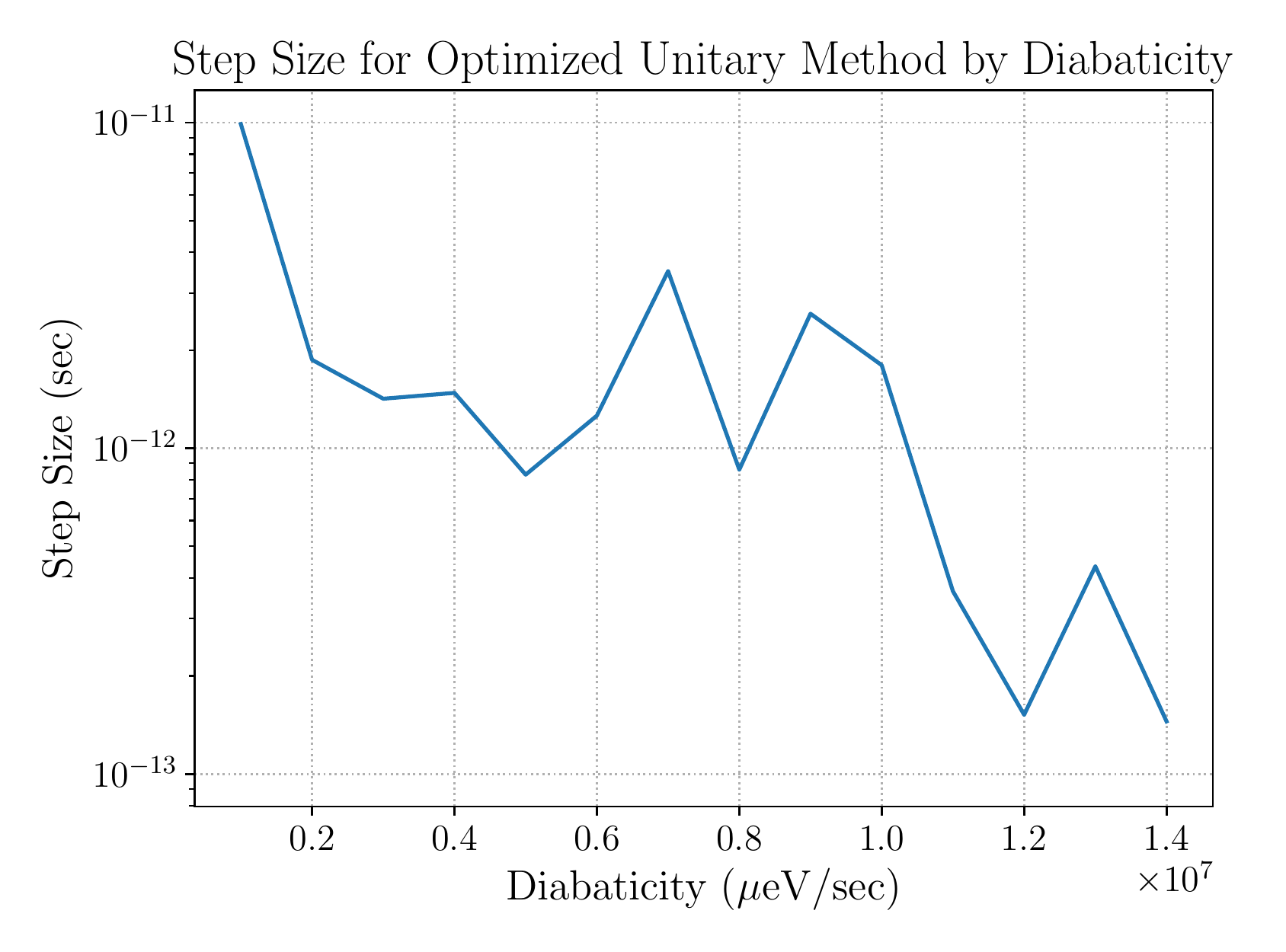}
			\caption{Step size required for the optimized unitary method to be accurate to within a tolerance of $1/64$, for states evolving with a three-state Hamiltonian and pulses from $-200~\mathrm{\mu eV}$ to $\{-100,0,100,\dots,1200\}~\mathrm{\mu eV}$ in $0.1~\mathrm{ns}$. For smaller diabaticities, larger step sizes are possible.}
			\label{fig:diabaticity}
		\end{figure}

		For rows, we calculated the step sizes required for an accuracy of $1/64$ for a variety of pulse heights (diabaticities); the results are plotted in Fig.~\ref{fig:diabaticity}. Within individual evolutions, adaptive step sizes may be implemented in the same spirit as the Runge-Kutta-Fehlberg Methods \cite{fehlberg1968classical}. We expect that adaptive step sizes could further accelerate computations by approximately one order of magnitude (while also increasing code complexity).
		
		Noting that each row of probability $P(\epsilon)$ is computed independently from each other row, the \verb|for| loop over detunings may be parallelized. This is expected to lead to performance gains in direct proportion to the degree of parallelization up to $\mathrm{num}$-fold improvement. Other performance considerations include the choice of programming language (including Cythonization) and computational hardware.
	
	\section{Implementation of Optimizations}
	\label{sec:appendix_pseudocode}
		Here we present pseudocode implementations for the simple unitary method, as well as for the three techniques of optimizing the unitary method as used for the square pulse, trapezoid pulse, and sine (or ramp) pulse.
		
		For the simple unitary method, applied to any detuning pulse, pseudocode would look like:
		\begin{verbatim}
		declare physical parameters
		for detunings:
		    for pulse times:
		        compute U in many steps
		        |psi> = U|psi_0>
		        further evolution and averaging
		        save data
		analyze and plot data
		\end{verbatim}
		
		For a square pulse composed of a plateau followed by a period of averaging, pseudocode would look like:
		\begin{verbatim}
		declare physical parameters
		compute U_readout
		for detunings:
		    for pulse times:
		        compute U = U_plateau in one step
		        |psi> = U|psi_0>
		        further evolution and averaging
		        save data
		analyze and plot data
		\end{verbatim}
		
		For a trapezoidal pulse composed of a rise, plateau, and fall, followed by a period of averaging, pseudocode would look like:
		\begin{verbatim}
		declare physical parameters
		compute U_readout
		for detunings:
		    compute U_rise and U_fall
		    for pulse times:
		        compute U_plateau in one step
		        U = U_fall U_plateau U_rise
		        |psi> = U|psi_0>
		        further evolution and averaging
		        save data
		analyze and plot data
		\end{verbatim}
		
		For a sine (or ramp) pulse composed of a uniformly rising ramp, followed by a period of averaging, pseudocode would look like:
		\begin{verbatim}
		declare physical parameters
		compute U_readout
		for detunings:
		    U = identity matrix
		    for pulse times:
		        compute U_new
		        U = U_new U
		        |psi> = U|psi_0>
		        further evolution and averaging
		        save data
		analyze and plot data
		\end{verbatim}
		
		Python scripts that implement the optimizes methods may be found in the supplementary material.
		
	\bibliographystyle{apsrev4-1}
	\bibliography{article}
	
\end{document}